\input epsf
\documentclass[nohyper,notoc]{article} 
\usepackage{epsfig}

\def\bit{\begin{itemize}}
\def\eit{\end{itemize}}
\def\ben{\begin{enumerate}}
\def\een{\end{enumerate}}
\def\beq{\begin{equation}}
\def\eeq{\end{equation}}
\def\bea{\begin{eqnarray}}
\def\eea{\end{eqnarray}}
\def\bq{\begin{quote}}
\def\eq{\end{quote}}
\def \lsim{\mathrel{\vcenter
     {\hbox{$<$}\nointerlineskip\hbox{$\sim$}}}}

\def\gappeq{\mathrel{\rlap {\raise.5ex\hbox{$>$}}
{\lower.5ex\hbox{$\sim$}}}}
\def\lappeq{\mathrel{\rlap{\raise.5ex\hbox{$<$}}
{\lower.5ex\hbox{$\sim$}}}}

\def\a{\alpha}

\begin{document}
\renewcommand{\thefootnote}{\fnsymbol{footnote}}
\begin{center}
{\Large {\bf 
Axions: Bose Einstein Condensate or Classical Field? }}
\vskip 25pt
{\bf   Sacha Davidson \footnote{E-mail address:
s.davidson@ipnl.in2p3.fr}  
} 
 
\vskip 10pt  
{\it IPNL, CNRS/IN2P3,  4 rue E. Fermi, 69622 Villeurbanne cedex, France; 
Universit\'e Lyon 1, Villeurbanne;
 Universit\'e de Lyon, F-69622, Lyon, France
}\\
\vskip 20pt
{\bf Abstract}
\end{center}

\begin{quotation}
  {\noindent\small 
The axion is a motivated dark matter candidate, so it would be 
interesting to find  features in Large Scale Structures
 specific to axion dark matter. Such 
features were proposed for  a Bose Einstein condensate of axions, leading to
confusion in the literature (to which I contributed) about whether  axions 
condense  due to their gravitational interactions. This note argues that
the Bose Einstein condensation of axions is a red herring: the
axion dark matter produced by the misalignment mechanism is
already a classical  field, which has the distinctive features attributed
to the axion condensate (BE condensates are described as classical fields). 
This note also estimates that the rate at 
which axion particles  condense  to the field, or
the field evaporates to particles,
is negligeable.

\vskip 10pt
\noindent
}

\end{quotation}

\vskip 20pt  

\setcounter{footnote}{0}
\renewcommand{\thefootnote}{\arabic{footnote}}


\section{Introduction}
\label{intro}

The axion  \cite{rev,PQ,revggr,DFS,russes} 
 is a light pseudo-goldstone  boson ($m_a \lsim m_\nu$), 
introduced  \cite{PQ} in
a solution of  the strong CP problem of QCD. It can
constitute the Cold Dark Matter(CDM) of the Universe. If the 
 phase transition at which axions appear occurs after
inflation, then  axions are unconstrained by observations of 
 inflationary density fluctuations({\it e.g.} PLANCK, BICEP2),  
and there are two
axion  contributions to CDM : oscillations of the classical
axion field  produced by the ``misalignment mechanism''
 \cite{DineFischler,PWW}, and
a population of non-relativistic modes  radiated by  strings
\cite{strings,DavisShellard}.
These two contributions can provide \cite{strings} 
the observed  CDM.

 Sikivie has raised the interesting question of whether axion dark matter 
could be distinguished from Weakly Interacting 
Massive Particles (WIMPs)\cite{DM}. This is pursued by 
various experiments which  search for  axions\cite{ADMX}
\footnote{see also\cite{EMR}.} and/or\cite{axion@WIMP}  WIMPs \cite{EdelCDMS,Xenon}. 
Sikivie and collaborators  \cite{SY1,ESTY}
noticed that the stress-energy tensors for axions 
and WIMPs are different, and 
  proposed that axion dark matter
could observably differ from WIMPs in non-linear structures.
For example, if  the  dark matter halo of a rotating galaxy were composed
of   a Bose Einstein (BE) condensate of axions, then
 vortices  could form, leading to  observable 
caustics in the dark matter distribution. This interesting
scenario has generated  discussion, both of  the
rate  at which the  Bose Einstein condensate  could form 
\cite{SY,SYetal2,DE,BJ},
and of the behaviour of a halo  of condensate\cite{RDS,BS13}.

The aim of this note is to argue that the issue of whether
axions are a  Bose Einstein condensate is a red herring.  
The Path Integral should allow to compute
anything, and in the Path Integral, CDM axions can be described
by the ``classical field'',  and the two-point function  (or equivalently,
the number density of particles). The dynamics should be controlled 
by the Lagrangian of a  scalar field coupled to gravity,
which is simple and well-known. I claim that practically,  
the classical  misalignment  axion field is  
always a Bose Einstein condensate (if one wishes to
use those words),
 and its evolution under gravity
 can be understood by solving  Einsteins Equations 
for a classical field \footnote{This means that  
discussions of the rate at which the 
misalignment field forms a BE condensate are irrelevant
--- despite the  considerable confusion (to which I contributed)
in the literature about the 
condensation of the misalignment field. }.
This is known to a segment of the community: Peebles studied
classical scalar field dark matter\cite{Peebles}, 
and 
in a beautiful series of papers, Rindler-Daller and Shapiro\cite{RDS}
study  whether it is energetically favourable for  a galactic
halo made of classical field to form vortices, and find
that  the $\phi^4$ coupling of the QCD  axion is of the wrong sign.

The outline of this paper is as follows.  The next section 
gives a brief review of axions and the BE condensate literature, and
 proposes a  Path-Integral-motivated
 translation dictionary among  the various vocabularies used to describe
cold dark matter axions. In particular, BE condensate 
= classical field.   Section \ref{sec:Tmn} 
reviews the stress-energy tensor  $T^{\mu \nu}$   for  the (axion)  field
and  for particles, because  
 $T^{\mu \nu}$ determines the evolution of dark matter, 
 in the classical  approximation. 
 Section \ref{sec:est} gives some rough estimates of
the ${\cal O} (G_N^2)$ rate at which gravity might transfer axions between
the particle bath and the classical field. This section,
estimating the rate at which axion
particles could condense to the classical field,
 is the only new part of this note.


\section{ Notation and notions }
\label{sec:notn}

This section contains a brief review of axion cosmology, 
an outline of some literature
on the BE condensation of axions, and proposes
  that axion CDM can be described
as a classical field and a distribution of cold particles.
More complete references can be found in \cite{DE},
a more thorough cosmological discussion in \cite{rev,WantzShellard}, 
astrophysical bounds in \cite{revggr}, and up-to-date
numbers in \cite{PDG}. This paper focuses on the QCD
axion; the more generic case of Weakly Interacting Slim Particles,
and Axion-Like Particles, are 
reviewed in \cite{WISP}.

\subsection{A brief review of axion cosmology}

In ``invisible'' axion    models \cite{DFS,russes},
the Standard Model is extended by a global ``Peccei Quinn''  U(1)
symmetry, and various new fields. This Peccei Quinn symmetry
breaks spontaneously at some high scale,  $f_{PQ} \sim 10^{11}$ GeV
for concreteness, where all the new fields
become massive  except the goldstone, who will become
the axion. 
 I suppose that this phase transition occurs
after inflation. 
This is consistent with the BICEP2 \cite{BICEP}
value of the inflationary expansion rate $H\sim 10^{14}$ GeV,
and avoids ``isocurvature bounds'' \cite{BGBL,PLANCK}
on axion CDM. After the Peccei-Quinn phase transition, 
the phase of the symmetry-breaking field, which
is the axion, 
takes an arbitrary value  between $-\pi$ and $\pi$ in each horizon volume,
and 
there is on average  a string per horizon (I suppose
a potential which does not allow more dangerous
defects\cite{autres}, such as domain walls).
Then the Universe expands until the QCD phase transition.
During this period, 
the coherence length of the  (massless) axion field grows with
the horizon \cite{BGBL}, and there remains on average a 
string per horizon\cite{strings}. As the QCD phase transition occurs,
the massive pion appears, and   mixes with  goldstone.  This
``tilts the mexican hat'', smoothly turning
on the axion mass.   Once $m_\pi$ reaches
a constant,  the axion field $\phi$ has 
a potential \cite{rev}
\beq
V(\phi) \approx f^2_{\rm PQ}m^2 [1-\cos(\phi/f_{\rm PQ})] \simeq \frac{1}{2} m^2 \phi^2
-  \frac{1}{4!} \frac{m^2 }{f_{\rm PQ}^2} \phi^4 
+  \frac{1}{6!} \frac{m^2 }{f_{\rm PQ}^4} \phi^6
~+ ...
\label{eqn3}
\eeq
where the axion mass is
\beq
m \simeq \frac{m_\pi f_\pi}{f_{\rm PQ}}  \frac{\sqrt{m_um_d}}{m_u+m_d}
 \simeq 6 \times 10^{-5} eV 
\frac{10^{11}  GeV}{f_{\rm PQ}} ~~~.
\label{mf}
\eeq

This potential has two implications for CDM axions: the
strings go away,  via  a complicated process studied
numerically in \cite{strings}, who obtain that
the energy in  the string network  is transfered to 
a population of  modes with momenta $\sim H_{QCD}
 \simeq 2\times 10^{-20} {\rm GeV } $. Once the
axion mass reaches its current value,  these
incoherent non-relativistic modes
contribute  \cite{strings} 
$
\Omega_a  \sim 0.2
\times 
\left({f_{\rm PQ}}/{ 10^{11} ~{\rm GeV}}\right)^{6/5}$.
The other effect of the potential (\ref{eqn3}), is to
cause the axion field $\phi$,  randomly located in
$(-\pi, \pi)$ in each horizon volume,
to roll towards its minimum, and oscillate. 
The   QCD horizon scale is  $H^{-1}_{QCD} $,
so the oscillations  of this ``misaligned'' 
classical axion field are non-relativistic,
and redshift like CDM\cite{aCDM}. They also grow large-scale
density fluctuations like CDM 
\cite{linearaxions,HwangNoh,ratra,NambuSasaki}.
The fluctuations in  the density of the
axion field are ${\cal O}(1)$ on the comoving scale
of  $H_{QCD} ^{-1}$ ($\sim 10^{-5}$ the distance
to the galactic centre today); the interesting
behaviour of these ``axion miniclusters''  is discussed in \cite{mini}.

\subsection{The scenario of axion  condensation}

The ``occupation number''  \footnote{Recall
that  the classical field and particle limits of a 
Lagrangian  require a different distribution of $\hbar$
\cite{hbar},
so that $\hbar$ is  explicitely required, to obtain
the particle number in a classical field configuration.}
of the misalignment field  is very high. In a seminal paper,
Sikivie and Yang\cite{SY1} proposed  that 
``gravitational thermalisation'' of the misalignment 
axions could cause them to Bose Einstein condense. They estimated
the gravitational interaction rate, or graviton exchange rate,
\beq
\Gamma_{int} \sim \frac{G_N m^2 n_\phi}{H^2}
\label{Gint}
\eeq
where $n_\phi$ is the axion number density.
This is faster than  the Hubble expansion rate $H$ for 
photon temperatures of $\lsim $ keV. This estimate,
for the gravitational interaction rate of the misalignment axions,
was confirmed in \cite{ESTY,SY,SYetal2,DE}. Notice that the
rate is linear in $G_N$, because the misalignement axions
are in a coherent state (classical field).  Sikivie and Yang then go on
to study the evolution of the axion  condensate using the
non-relativistic Schrodinger equation (or Gross-Pitaevskii equation),
as used later by Rindler-Daller and Shapiro\cite{RDS}. 

 The rigourous
analyses of Saikawa et al  \cite{SY,SYetal2} focus on
the gravitational interactions of misalignement axions, and
obtain equations of 
motion for the axion number operator in second quantised field theory, 
both in flat  Newtonian space-time, and 
in  a perturbed expanding Friedmann-Robertson-Walker Universe.
They  show that $dn_k/dt  \simeq \Gamma_{int}$, where $n_k$
is the number of axions of momentum $k$ in the coherent
state representing the misalignment axions.

In a previous paper with Elmer,  we doubted that 
(\ref{Gint}) was a thermalisation rate, because there
is no entropy generation (no fluctuations are averaged over,
the evolution is coherent).  Using the equations of motion 
for a classical field in an expanding Universe with small
density fluctuations, we reproduced eqn (\ref{Gint}). 
We also  recalled  that, with a different parametrisation,
these equations are linear and can be solved: if one studies
the fourier modes of the energy density, rather
than those of the axion field,  then one obtains the familiar
equations for density fluctuations in the early Universe.
We interpreted that the gravitons exchanged 
in  eqn (\ref{Gint})  are  coherently
growing density fluctuations in the early Universe. 

Recently, Berges and Jaeckel\cite{BJ} recall that thermalisation
is not required  for Bose Einstein condensation, and make
analogies to results obtained in $\lambda \phi^4$ theories. 
They recall that an initially highly occupied distribution of
low-momentum particles  develops an infrared cascade,  and
that a condensate can form  before the high energy tail
of the distribution has reached thermal form.  Applied to
CDM axions,  the observation of \cite{BJ} could suggest  that
the cold axion particles produced by strings,  might
join in the axions in the misalignment field. I return
to this in section \ref{sec:est}.

\subsection{Proposals from the path integral}

Whether axion-CDM 
behaves differently from WIMP-CDM, is an interesting question.
To identify variables and equations with
which to address it,    one can consult the path integral,
which in principle knows everything.

The Path Integral  allows to compute $n$-point functions, 
in particular,  
the expectation value of the field $\langle \phi (x,t) \rangle =
\phi_{cl}(x,t)$,
and of the two point function.
  The ``classical field''  $\phi_{cl}(x,t)$ is familiar from the 1PI
effective action. In closed time path formalism, the
two point function encodes a statistical number (or phase space)
density, as well as the propagator.  Since axions are
feebly interacting, the higher point functions 
can  often be neglected.  
The classical field and the density 
of incoherent modes are  convenient variables for the discussion
of cold Dark Matter axions, precisely  they can
be identified as  the misalignment field,
and the modes radiated by strings.

The classical field is distinguished from
 a distribution of  particles by its  macroscopic 
coherence. It can be represented in second-quantised Field Theory
as a coherent state \cite{I+Z}, which is constructed by
acting on the vacuum with the exponential of the creation
operator (see eqn (\ref{etatcoh})). Such a  state
is therefore {\it not} an eigenstate of the number
operator, instead, the expectation value of the
field operator in the coherent state gives the
classical field.   It follows that the
classical field is something like an amplitude, 
and its  equations of motion 
can be linear in
the coupling. This differs from  particles,
where only forward scattering is linear in the coupling,
because it interferes with doing nothing.

The statistical part of the two-point function describes an
incoherent distribution of modes or particles. This
paper assumes that axion strings decay into such
a  distribution
\footnote{ We plan to  address this
question in more detail in a subsequent publication. }.
Since the time evolution of a statistical distribution
of classical modes, or particles, is approximately the
same \cite{AB}, for concreteness, it is supposed
here that the strings decay into cold  axion particles.

The path integral can also provide equations of motion for
the classical field and number density, although
more familiar is the  recipe to
obtain in-out  vaccuum $n$-point functions,
with which  to calculate cross-sections. However,
 initial value problems \cite{Mot}  can   
can be posed  in the Path Integral,  
using   the Schwinger-Keldysh or 
Closed Time Path formalism \cite{Schwinger}.
Furthermore,  in the 2PI  formalism\cite{CJT}, the
classical field and the two point function 
appear as variables.  So the indicated formalism
for studying axion CDM would be the Closed Time
Path 2PI action for axions, prefereably in curved 
space-time. An $O(N)$   $\lambda \phi^4$  theory
was studied in this formalism in \cite{RamseyHu}.
However, this note uses more simple
and familiar formalism.  Einsteins equations and $T^{\mu \nu}_{~~;\nu} = 0$
are used  as equations of motion, and 
the stress-energy tensor  of the axion  field
and particles is evaluated 
as  the expectation value of an operator
in the usual  way (``in-out vacuua''). 
The classical equations of motion  should
be  acceptable because gravity is a classical
theory and the axion is feebly coupled, and
equating in and out vaccua should be acceptable
again because the axion is feebly coupled.

\subsection{What is a Bose Einstein condensate of axions?}
\label{sec:dic}

The axion literature uses diverse vocabulary and  calculational
techniques. I propose to assume the following translation dictionary
\beq 
{\rm classical~ field~ =~ condensed~ regime~ =~ Bose~ Einstein~ condensate~~.}
\label{dict}
\eeq
That is, the misalignment axions are a Bose Einstein condensate. 
And eqn (\ref{Gint}) is irrelevant, because it is the
gravitational interaction rate of the
axions in the condensate.

{What is a Bose Einstein condensate?}
Bogoliubiov \cite{Bog}
 long ago identified the Bose Einstein condensate as the
macroscopic  occupation of   the lowest energy mode. In particular,
starting from a second-quantised formalism, he treated  as numbers the 
creation and annihilation operators of the zero mode\footnote{This is
related to the coherent state, eigenstate of the annihilation operator.}, 
such that
the field operator could be written as a  classical field in
the zero mode plus creation and annihilation operators for the
remainder of excitations. This formalism was used by
Nambu and Sasaki to describe density fluctuations
in the axion misalignment field\cite{NambuSasaki}.

Important characteristics of a  BE condensate seem to be 
\ben
\item a classical field, 
\item carrying
a conserved charge,  
\item whose fourier modes are
concentrated at a particular value --- that is, most
of the ``particles''
who condense to the  BE condensate, should coherently be
doing the same thing. However, they
do not need to be in the zero-momentum mode.
\een  This is consistent with 
Bose Einstein condensation in equilibrium statistical
mechanics and   finite temperature field theory\cite{KapustaBEC,HW}
(where in homogeneous systems, the bosons condense
in the zero-momentum mode),  as well
as with  the experimental studies of BE condensation in
alkali gases\cite{Leggett,PitRev}, where the condensed atoms  have
similar velocities.

Are the misalignment axions  a BE condensate?
The axion is a real (pseudo)scalar (so formally has no
conserved charge), but the number changing interactions
are sufficiently slow that axion number is approximately conserved.
 If the PQ phase transition
is after inflation, the fourier  modes
have an approximately white noise spectrum,
so the classical axion field is not peaked
at a particular momentum mode. 
However, readers with a predilection for
BE condensates, who are attached to the
third criteria,   could then view the axion field  as a superposition
of BE condensates, which are coupled via gravity \cite{DE}.

In any case, the question of whether the axion
field is a BE condensate seems more about vocabulary 
than dynamics. Axion cold dark matter
should evolve according to its equations of motion,
which  are
approximately
those of a (free)  non-relativistic  scalar interacting with gravity.
These are the equations studied in \cite{SY1,RDS}.
In my opinion, the BE condensate analogy is
not useful for axions, because the  familiar BE condensates
($^4$He, etc) have  stronger  short range interactions
than axions.


\section{Formalism to calculate with: stress energy tensors}
\label{sec:Tmn}

The matter current which couples to 
gravity  is  the stress-energy tensor, so this section reviews
the stress-energy tensor of a (non-relativistic)
 classical scalar field (the misalignment
axion field), and of a  distribution of particles(the cold axions
from strings).    The  equations  
which govern the formation of galaxies and large structures  can
then be respectively obtained from 
$T^{0\mu}_{~;\mu} = 0$ and
$T^{i\mu}_{~;\mu} = 0$. 

Recall that the stress-energy tensor for dust, or 
non-relativistic  non-interacting
particles with $U^\mu = (1,\vec{v})$,  is
\beq
T_{\mu \nu} = \rho U_\mu U_\nu =
\left[
\begin{array}{cccc}
\rho & &-\rho\vec{v}&\\
&&&\\
-\rho\vec{v}&& \rho v^iv^j&\\
&&&
\end{array}
\right] ~~~ ~~~({\rm dust}).
\label{Tmndust}
\eeq
With a  metric\footnote{Alternatively, the gravitational interaction can be put ``by hand''
into the Euler equation in Minkowski space.}
\bea
ds^2 & = & (1 + 2\psi)dt^2 - (1 - 2 \psi) \delta_{ij} dx^i dx^j ~~~,
\label{metrique}
\eea
 which
 describes Minkowski space with a Newtonian
potential  $\psi$ (satifying
 $\nabla^2 \psi = 4\pi G_N \rho$  by
Einsteins Equations), 
then  $T^{\nu \mu}_{~;\mu} = 0$ gives
\bea
\partial_t \rho + \nabla\cdot (\rho \vec{v}) = 0 && {\rm continuity}
\label{cont}
\\
\partial_t \vec{v}  +
( \vec{v}\cdot \nabla)  \vec{v} = -\nabla \psi   && {\rm Euler}~~~,
\label{Eul}
\eea
which should describe   the evolution of
a galactic halo made of cold  non-interacting particles \cite{efstathiou}.

To obtain the stress-energy tensor
for the axion field and particles, 
I work
 in second-quantised field theory, in the
Heisenberg picture (time dependent operators).
The axion is real
and non-relativistic, however, 
it is convenient to use  covariant (relativistic) 
 notation $x^\a = (t,\vec{x})$ 
and a complex field $\phi$. The relativistic
notation  is
because  $T^{\mu \nu}$ is covariant. 
Then, the almost-non-interacting non-relativistic
axion field has an effectively conserved quantum
number (particle number). Indeed, a real
relativistic field $\varphi = \varphi^\dagger$ can
be described in the non-relativistic limit by
a complex field $\phi$:
$$
\varphi = 
\phi e^{-imt} + \phi^\dagger e^{imt}
~~.
$$
So to have a conserved number  in relativistic notation,
I use a   complex scalar field,  which has a
conserved  current.  It is straightforward to check that
the stress-energy tensors obtained for the field
and the cold particles (the anti-particle modes
are neglected) will be the same.
In this note, the non-relativistic limit
is taken by neglecting $\partial^2_t$ and  $(\partial_t ~)^2$, 
and by neglecting  $\partial_t \theta$ (see
 eqn (\ref{defntheta})) with respect to $m$.

 The field operator
can be  fourier-expanded 
\beq
\hat{\phi} (x) = \int \frac{d^3 k}{(2\pi)^3}
\frac{1}{\sqrt{2E_k}}\left( \hat{a}_k e^{-ik\cdot x}
+ \hat{b}_k^\dagger e^{ik\cdot x} \right)
\label{phi}
\eeq
 on particle annihilation and  anti-particle creation  operators satisfying
$[\hat{a}_k, \hat{a}_p^\dagger] = \delta^3(\vec{k} - \vec{p}) (2 \pi)^3$.

The    axion misalignment
field can be represented as a coherent state\cite{I+Z}:
\beq
|\phi \rangle = \frac{1}{N} \exp\left\{ 
\int \frac{d^3 q}{(2\pi)^3}
\widetilde{\phi}(\vec{q})
 \hat{a}^\dagger_q
\right\} |0\rangle
\label{etatcoh}
\eeq
where $N$ is a normalisation factor such that $
\langle \phi |\phi \rangle = 1$. The coherent
state is constructed such that the
expectation value of the field operator is the classical field $\phi(x)$:
\beq
\langle \phi |\hat{\phi} (x) |\phi \rangle
= \int \frac{d^3 k}{(2\pi)^3}
\frac{1}{\sqrt{2E_k}} \widetilde{\phi}(\vec{k}) e^{-ik\cdot x}
=\phi (x) = \eta(x) e^{-i\theta(x)} =  \eta(x) e^{-i(mt + S(x))} 
~~~.
\label{defntheta}
\eeq
The fourier mode expansion of $\phi$  is convenient
for linear perturbations in the early Universe
\cite{DE}, and 
for evaluating expectation values in  a coherent state.
However, the parametrisation
 $\phi = \eta e^{-i\theta} $
(with $\eta$ and $\theta$  real)
is more appropriate for  a classical field
with a  conserved  number (a Bose Einstein condensate?),
 and $\theta = mt + S$ will be  useful  in 
the non-relativistic limit.  A similar
parametrisation is used  by  Daller-Rindler
and Shapiro \cite{RDS}.

\subsection{The classical field}

  The stress-energy tensor for a complex scalar 
 $\phi = \eta e^{-i\theta}$,   with potential
$V(\phi^\dagger \phi) = m^2\phi^\dagger \phi + 
\lambda (\phi^\dagger \phi)^2 $ is
\bea
T_{\mu\nu}&= &\partial_\mu \phi^\dagger    \partial_\nu \phi+ 
\partial_\nu \phi^\dagger    \partial_\mu \phi
-  g_{\mu \nu} \left( \partial^\a \phi^\dagger    \partial_\a \phi
 - V(\phi^\dagger \phi) \right) ~~~~~~{\rm (classical~ field)}
\nonumber \\
&=& 2 \partial_\mu \eta    \partial_\nu \eta+ 
2\eta^2 \partial_\nu \theta     \partial_\mu \theta
-  g_{\mu \nu} \left( \partial^\a \eta    \partial_\a \eta
+ \eta^2 \partial_\a \theta     \partial^\a \theta
 - V(\eta^2) \right) ~~~,
\label{TmnGR}
\eea
where $T_{00} = \rho$, and $T_{ii} = P$ (no sum on $i$). 

The flat space  equations of motion (gravity will be included
by hand) are
\bea
0 = \partial^\mu T_{\mu \nu} &=& \partial_\nu\eta
\left\{  2 \partial_\mu     \partial^\mu \eta - 
2\eta  \partial_\a \theta     \partial^\a \theta
+\frac{\partial V}{\partial \eta}\right\} 
+  \partial_\nu\theta
{\Big \{}\partial^\mu
\left(\eta^2 \partial_\mu \theta 
 \right) {\Big \}}
\label{EdM}
\eea
where in curly brackets are the  equations of motion that
would be obtained from the Lagrangian.
In particular, writing
$\theta = mt + S$  
 in the non-relativistic limit,
the current conservation
equation 
\bea
\partial^\mu (\eta^2 \partial_\mu \theta) &\simeq &
\partial_t {\Big (} m \eta^2  {\Big )} -
\partial_j  {\Big (}  \eta^2 \partial_j S  {\Big )}  
\label{conserv}
\eea
 becomes the continuity
equation $\partial ^\mu T_{\mu 0}  = 0$, 
with the approximations
\bea
T_{00}= \rho  &= & 2m^2 \eta^2 + ...\\
T_{j0}  &=&  2m  \eta^2 \partial_j S +... ~~~.
\eea
where  ``$+...$'' can represent derivatives of $\eta$ and
$S$ and the $\lambda \eta^4$ interactions.
With the identification ${v}^j = -\partial_j S/m$, 
 $\partial ^\mu T_{\mu 0}  = 0$ for the field 
 is  identical to  eqn (\ref{cont})  for cold, non-interacting particles.

As is well-known \cite{Peebles,SY1,RDS},   a classical field has 
 additional contributions to $T_{ij}$ with respect to
(\ref{Tmndust}):
\bea
T_{ij} &=& 2 \partial_i \eta    \partial_j \eta+ 
\rho v_i  v_j
+  \delta_{ij} \left(  -\nabla \eta  \cdot  \nabla \eta
- \rho |\vec{v}|^2 + 2m\eta^2 \partial_t S  
 - \lambda \eta^4 \right) ~~~,
\label{Tij}
\eea
where $\lambda = -m_a^2/(12 f_{PQ}^2)$ from
eq. (\ref{eqn3}).
However, 
 the equations of motion,
given in \cite{RDS}, are 
more useful for understanding the
modified dynamics. From the first equation in
curly brackets of (\ref{EdM})\footnote{Alternatively,
$\partial^\mu T_{\mu j}=0$, combined with
current conservation,  gives an equation for
$\partial_t {S}$. Taking its gradient gives  the Euler-like 
eqn (\ref{extra}).}, an Euler-like equation can be  obtained:
\bea
\rho \partial_t \vec{v} + \rho ( \vec{v}\cdot \nabla)
\vec{v} = -\rho \nabla \psi  -\rho\nabla Q   -\nabla  P_{SI}
\label{extra}
\eea
where $Q = -\frac{1}{ 2m \eta} \nabla^2 \eta$  describes
the ``quantum kinetic energy'' or
``quantum pressure''\cite{PitRev}  of the classical field, and
$P_{SI} =  2\lambda \eta^4$  is proportional to the pressure arising
from the  Self Interactions of the field. 
Comparing to (\ref{Eul}), shows that the classical scalar
field has extra forces \cite{Peebles,SY,RDS} compared to dust. 

As noted by Rindler-Daller and Shapiro, the sign of 
$P_{SI}$ is important:  negative pressures and potential
energies (such as the gravitational $\psi$) induce  attractive
forces, whereas positive $P_{SI}$ would give a repulsive
force, counter-acting gravity.  The axion potential of eqn (\ref{eqn3})
gives a negative $\lambda\phi^4$ coupling, so the
self-interactions of axions induce an attractive force
which encourages the field to clump. This is a well-known
behaviour for BE condensates in atomic traps \cite{PitRev}.
Rindler-Daller and Shapiro  look for
stationary rotating solutions to eqn (\ref{extra}),
which contain a vortex. They find that the ``quantum pressure''
of a scalar field as massive as the axion is insufficient
for a vortex to be  energetically favourable.  And since
the $\lambda\phi^4$ coupling of axions is negative,
the resulting  force  attracts axions
towards the centre of the halo, which also discourages
vortices \cite{RDS}.

\subsection{Cold particles}

First,  its worth to check that the usual WIMP-CDM
eqns apply   for the  cold axion
particles. This requires obtaining the ``classical
particle limit'' from second-quantised field theory,
which should be possible using Wigner transforms.

The stress-energy
tensor   can be
introduced  as a function  
of two space-time points $T_{\mu\nu}(x_1,x_2)$,
separated by a small distance $\delta$. 
Then performing
a Wigner transform\cite{Wigner}
\beq
T_{\mu \nu}(X,Q) =
\int \frac{d^4 \delta}{(2\pi)^4}e^{iQ \cdot \delta} 
T_{\mu \nu}(X-\delta/2,X+\delta/2)~~~,
\label{wigner}
\eeq
allows to obtain a (classical) distribution
of particles of three-momentum $\vec{Q}$, at point $X$.  This 
classical approximation is expected to be acceptable
because there is a separation of scales between 
the (galactic) distances parametrised by ${\vec{X}}$,
and the inverse-axion-momentum scale $|\vec{\delta}|$
(the inverse-momentum of an axion of  $10^{-4}$ eV with $v=300 $km/s 
is a few metres). Intuitively,
the coordinate $\vec{X}$ can be imagined  discrete, so the
galaxy is parametrised by a grid, with
 a cube in which particles are quantised
 at each point
of the grid. 
Therefore, the creation and annihilation operators 
are also labelled by  $X$, and 
 the multi-particle state in which 
$\langle T_{\mu \nu}(X,Q)\rangle $ is evaluated
can have a  different number distribution of
particles at each point.

The  quantum field theory operator representing the density can be
written (in the $\lambda\to 0$ limit)
\bea
\hat{\rho}(X,Q) = \int \frac{d^3k}{(2\pi)^3}  \frac{d^3p}{(2\pi)^3}
\frac{1}{\sqrt{2k_0 2p_0}}
{\Big (}  k_0 p_0 + \vec{p}\cdot \vec{k} + m^2 {\Big )}
\hat{a}^\dagger_k(X)   \hat{a}_p(X)
\frac{1}{(2\pi)^4}\delta^4(Q-\frac{p+k}{2}) e^{i(p-k) \cdot X}~~~.
   \eea 
Evaluating $\hat{\rho}(X,Q)$ in a multi-particle state
$| n\rangle$, defined such that 
$\langle  n |a_k(X)^\dagger a_p(X) | n \rangle =
f(X,k)\delta^3(\vec{k} -\vec{p}) (2\pi)^3
$, where
$f(X,k)$ is the distribution in phase space of axion particles,
and  integrating over
$Q$  (this reestablishes the correct mass dimensions  for $T_{\mu \nu}$, 
and removes a leftover  $\delta(p^2 - m^2)$), gives
\bea
{\rho}(X) = \int \frac{d^3q}{(2\pi)^3}  q_0
f(X,q) ~~~,
\label{rhopart}
 \eea 
which is the expected ``classical  particle'' result. In the non-relativistic limit, $q_0\simeq m$, and $\rho(X) = mn(X)$.

The energy flux $T_{0i}(X-\delta/2,X+\delta/2)$
can be manipulated in a similar manner. If I neglect
$\partial_t \hat{a}_k(X)$ with respect to $m \hat{a}_k(X)$
(the non-relativistic approximation),  and $\partial_i 
 \hat{a}^\dagger_k(X)$  with respect to  
$k_i \hat{a}^\dagger_k(X) $ (the separation of scales
$\delta \ll X$ discussed after eqn (\ref{wigner})), then
 \bea
T^{0i}(X) =  \int \frac{d^3q}{(2\pi)^3}   q^i
f(X,q)   ~~~,
\label{T0ipart}
 \eea 
(as given for instance in \cite{MB}).  Defining ${v}^i =
T^{0i}/\rho$, this reproduces the top row of eqn(\ref{Tmndust}). 


\subsection{What to do with this formalism?}

More interesting than
the  stress-energy tensors for axions, or their  equations of motion,
would be the solutions of the equations of motion. 
These are well-known in the period of linear structure growth,
where  the axion particles and field grow fluctuations
as do  WIMPs, because pressure is neglected in the
equations of motion on the relevant scales \cite{BGBL,DE}. 
It is interesting to wonder 
whether the extra pressures of the classical axion field,
as compared to axion particles or WIMPs,  could be
relevant during non-linear structure formation.
Structure growth in the non-linear epoch  can be studied numerically,
so this could be addressed by developing
 a code where Dark Matter  is evolved as a fluid. The 
pressure and viscosity of
the axion field could then be included, allowing to 
 identify  observable consequences of axion {\it field}  dynamics,
distinguishing the misalignment axions  from WIMPs. 
For such a code,  it would be important to know
the rate at which axions pass between the field and
the cold particle bath; this is estimated in
the next section.


\section{Gravitational evaporation of the field?}
\label{sec:est}

 This section estimates the rate at which
axions can transfer between the field and  the bath of particles. 
This is
interesting because the axion particles should behave
like WIMPs (both have the stress-energy tensor
of ``dust''), whereas the field could cause structure to
grow differently, due to its additional $T_{ij}$. 

First,  the rate mediated by graviton
exchange is estimated, because it is infrared divergent,
so could be large. Then I compare
to the axion self-interaction rate.     There are various
issues to address for such an estimate: the order in $G_N$,
the cutoff for  the infrared  divergent  graviton
exchange, the bose enhancement factors  from the high
occupation number of axions, and how to include
the axions from the classical field. They will be
addressed in that order.

\begin{figure}[ht]
\begin{center}
\epsfig{file=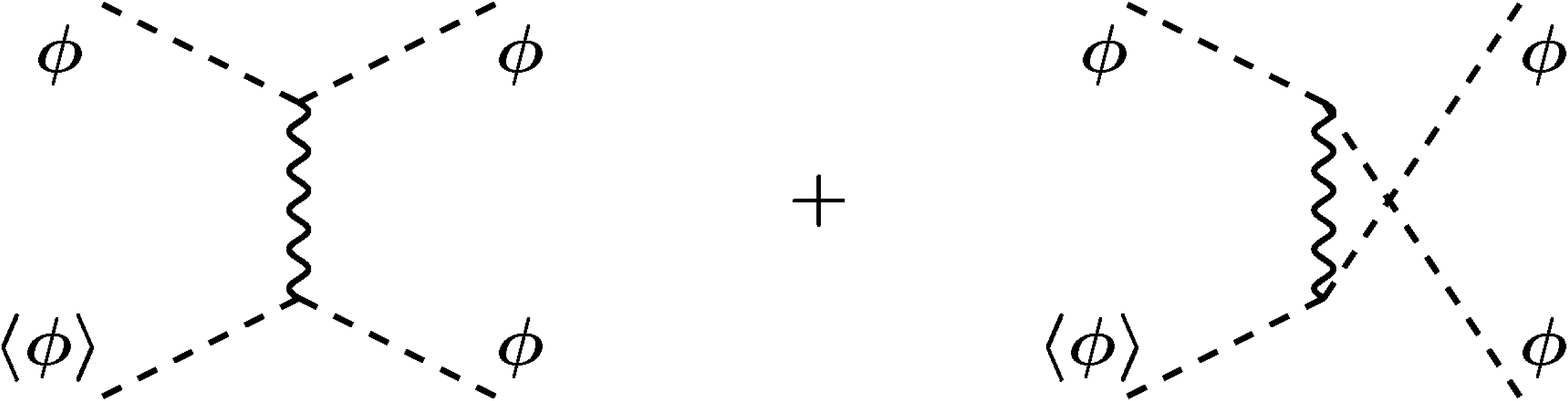, height=3cm,width=10cm}
\end{center}
\caption{
Feynman diagrams for  the gravitational scattering of
an axion from  the condensate $\langle \phi \rangle$ 
 with an axion from the bath,
resulting in two axions in the bath.}
\label{fig}
\end{figure}

The   estimate will turn out to
be  ${\cal O}(G_N^2)$, corresponding to  a cross-section
representing ``quantum'' graviton
exchange.  I imagine that it is reasonable to estimate quantum gravity
corrections using field theory at scales where field theory  is experimentally
verified, despite that  possible  destabilisations of the axion potential
from the Planck scale \cite{planck} are not discussed.

I assume that at ${\cal O} (G_N)$, the interactions between gravity
and matter are given by Einsteins Equations, where 
the stress-energy tensor is the  matter-current to which gravity
couples. 
As seen in the previous section,
evaluating the expectation value of  the operator
corresponding to the  stress-energy, in a state  composed
of a bath of axion particles and  a classical field, does
not give any ``cross-terms'', in which appear the 
the field and the particles. This is because any
interaction between the field and the
particles would change the number of particles, 
so the expectation value is zero  in a particular state
(corresponding to a distribution of
particles and a classical field).
Therefore at  ${\cal O} (G_N)$, the
misalignment field and the particles 
feel  each others stress-energy, but do
not exchange axions.

At ${\cal O} (G_N^2)$, the  cross-section for 
 the gravitational scattering of
non-relativistic bosons (see figure \ref{fig})
  is given  by DeWitt  \cite{DeWitt} as:
\beq
\frac{d \sigma}{d\Omega}  =
\frac{G_N^2m^2}{16}\left(
\frac{1}{v^2 \sin^2 (\theta/2)} +
\frac{1}{v^2 \cos^2 (\theta/2)} 
\right)^2 
\label{dewitt}
\eeq
where $v$  is the three-velocity of an incident axion in the
centre-of-mass frame, and I removed the
annihilation contribution included  in\cite{DeWitt}.  
The cross-section is clearly infra-red
divergent for small angle scattering, corresponding to
soft gravition exchange. If  one of the axion  legs
  is in  the coherent state representing
the  misalignment field,  then  the cross-section
describes the passage of axions between the  particle bath
and the field. That is, it could be involved in the
condensation of the particles to the field, or in the
evaporation of the field to particles. 

To determine the rate at which gravitational scattering
moves axions between the field and the bath, it is clear
that one must identify an infrared cutoff
for (\ref{dewitt}). I claim that it should be
$|\vec{\delta}|^{-1}$,  of order the axion
three-momentum, because   $|\vec{\delta}|$ was the
spatial scale in  the
Wigner transform, below which there were particles. 
That is, gravitons couple to  stress-energy
(in an almost-flat space),
because the stress-energy is the
variation of the action with respect to the metric. 
Concretely,  
the matrix element for gravitational scattering
of scalars given by DeWitt\cite{DeWitt} is:
\bea
{\cal M}[\phi(p_1) + \phi(p_2) \to \phi(p_3)+\phi(p_4)]  = 16 \pi G_N
 T^{\mu\nu}(p_1,p_3)
\frac{
g_{\mu \sigma} g_{\nu\tau} +g_{\mu \tau} g_{\nu\sigma} -
g_{\mu \nu} g_{\sigma\tau}
}{(p_1 -p_3)^2}  T^{\sigma\tau}(p_2,p_4) + u~{\rm channel}
~~~,
\label{M}
\eea
where $ T^{\mu\nu}(p_1,p_3) = 
\sqrt{\frac{1}{4E_1E_2}}[p_1^\mu  p_2^\nu + p_2^\mu  p_1^\nu 
-g^{\mu \nu} (p_1\cdot p_2 + m^2)]$  reduces to 
$ m \delta^{\mu 0} \delta^{\nu 0}$ in the non-relativistic limit. 
I claim that gravitons  interact with individual axions,
if the graviton momentum  allows to see inside the box
of volume $|\vec{\delta}|^3$.  On longer wavelengths, the
graviton sees the stress-energy tensor of the  particle distribution discussed
in the previous section, and cannot be taken to have
incoherent  interactions with individual  axions. That is,
the longer-wavelength graviton interacts coherently with
many axions, so these interactions must be summed in the amplitude,
giving rise to  graviton interactions with the stress-energy.  
 Therefore the infrared cutoff of the
${\cal O}(G_N)^2$ cross-section, which represents incoherent
graviton-axion interactions,  should be  the axion
3-momentum $\sim m \times 10^{-3} c$, which gives
\beq
\sigma_G \sim 10^4 \frac{m^2}{m_{pl}^4}
\label{grav}
\eeq
This  cutoff is intuitive, because we expect  a graviton of
wavelength the size of the earth, to interact
coherently with the earth, and not incoherently
with the individual gluons and quarks which make it up.
This also agrees with the scattering  of an MeV
photon on  a proton, where the photon sees
a point particle of charge +1, and not the
quarks inside. 

This cross-section can be compared to the rate
at which an axion particle could scatter an axion out of the
condensate via its $\lambda \phi^4$  coupling\footnote{Preskill,
Wise and Wilcek\cite{PWW} estimate the rate at which a condensate evaporates
via the six-axion coupling, in the process
[four condensate axions]$\to$[two axion particles]. This
is the lowest order kinematically allowed diagram for
the case they consider, of a condensate made of zero-momentum axions in
vacuum. If there are axion particles in the
initial state, as considered here, they can scatter axions out of
the condensate via the $\lambda\phi^4$ coupling.} 
\beq
\sigma_\lambda \simeq \frac{\lambda^2}{4\pi m^2} \sim
\frac{m^2}{4\pi (4!)^2f_{PQ}^4}
\label{sigmalambda}
\eeq
 which  exceeds eqn  (\ref{grav}) for $f_{PQ}\lsim 10^{-2}m_{pl}$.
If  the PQ phase transition
is after inflation (as supposed here),  then $ \sigma_\lambda > \sigma_G $.

The high occupation numbers
of the axion field (coherent state) and particles must 
be taken into account.  In the familiar Boltzmann
equation which would describe scatterings  
among axion particles, the rate at which  the number
density  $n$ of axion particles changes, includes
 axions being  scattered in minus
out of each  mode, so can be  written 
\bea
\frac{\partial }{\partial t}n  &=&
\int \frac{d^3p_1}{2E_1(2\pi)^3}
\frac{d^3p_2}{2E_2(2\pi)^3}
\frac{d^3p_3}{2E_3(2\pi)^3}
\frac{d^3p_4}{2E_4(2\pi)^3}
(2\pi)^4 \delta^4(p_1 + p_2 - p_3-p_4) \nonumber\\
&&~~ \times |{\cal M}[\phi(p_1) + \phi(p_2) \to \phi(p_3)+\phi(p_4)]|^2 {\Big [}
f_1f_2 (1+f_3)(1+f_4) - f_3f_4 (1+f_1)(1+f_2)  {\Big ]}~~. 
\eea 
It is clear that $f_1f_2 (1+f_3)(1+f_4) - f_3f_4 (1+f_1)(1+f_2) 
\sim f^3$. So the rate at which  axions move between the
bath and the condensate due to gravity is proportional
to the density of targets, multiplied by one
bose enhancement factor $f$, and can be estimated as
\beq
\Gamma \sim  n_\phi \sigma_G f \sim  10^{13} 
\left(\frac{\rho_{DM}}{\rho_c}\right)^2
\left(\frac{H_0}{m}\right)^3 ~ H_0
\label{neg}
\eeq
where $n_\phi \sim m \eta^2$ is the number density of axions
 in the  field, and   $f\sim n_{\phi}/(m^3 v^3)$ is the occupation
number of axion particles of non-relativistic velocity  $\lsim v$
(this estimate applies when the field and particles make
similar contributions to the dark matter densty).  The second estimate
in eqn (\ref{neg})  is  for the galaxy today,  using  $\rho_{DM}\sim mn_\phi
\simeq $ 0.3 GeV/cm$^3$, 
$v\sim 10^{-3}c$,  and that the
Hubble  rate today  is $H_0 \sim  \sqrt{\rho_c}/m_{pl}$  where
$\rho_c$
is the critical density.
With the  amusing  numerical coincidence
that $H_0 \sim  m^2/m_{pl}$, one sees that 
this rate is   negligeable, compared to 
the  expansion rate of our Universe today,
because  $H_0/m \sim m/m_{pl} \lsim 10^{-24}$.
Therefore
gravity does not move axions between the field and
particle bath, within the age of the Universe.

A similar estimate, using the cross-section
for axion self-interactions given in eqn (\ref{sigmalambda}),
would give a rate amplified by a factor $\sim (\frac{m_{pl}}{4\pi f_{PQ}})^3$.
This clearly cannot compensate the  $\sim (m/m_{pl})^3$ factor,
so it appears than neither gravity nor self-interactions
can move axions between the field and bath
within the age of the Universe.

Finally, in the above estimates, the axions in the coherent
state were treated  ``like particles'', except with a different
formula for the number density.    This should be reasonable,
as can be seen from the more correct analysis of \cite{NZG}.

\section{Summary}
\label{sec:sum}

The ``classical  field'', 
and the ``density of cold  particles'',
are understandable phrases  with which to describe
the two contributions which axions
can make  to dark matter. Avoiding
the issue of ``Bose Einstein Condensation'' 
allows to focus on the 
interesting question of
how to distinguish axions from WIMPs.  
Various studies\cite{Peebles,RDS} 
suggest  that  during non-linear structure formation,   
cold particles and a classical field could
 grow  galaxies differently,
due to the extra pressures and viscosities of 
the field.   
For instance, the analytic analysis of 
 Rindler-Daller and Shapiro (RDS)\cite{RDS},
includes a scalar field with
the parameters of   the  QCD axion. 
RDS  did not confirm that   vortices
in the halo are  a signature
of an axion field, but other observable
differences could be identified by  following the
dynamics of galaxy formation (RDS 
 look for stable solutions representing the galaxy today).
 So it would be
interesting to  model galaxy formation 
in the presence of scalar field dark matter, or
more generally, to study non-linear structure
formation with a code evolving dark matter as a
fluid with pressure and viscosities.  

For the case of axions, which can contribute
two components to dark matter, it is relevant to
estimate the rate at which axions could move between
the field (distinguishable from WIMPs), and
the cold particle bath. This note made simple
estimates for   the evaporation/condensation rate
of the field in the presence of a bath of axion
particles,  due to   gravitational interactions,
or the $\lambda \phi^4$ self-interactions. Both
rates were found to be negligeable compared to
the Hubble expansion rate, suggesting that
the axion field remains coherent during
the violent process of galaxy formation.

\section*{Acknowledgements}

I thank the Max Plank Institut f\"ur Physik and
SISSA for seminar invitations,
  F Br\"ummer, A de Simone, M Nemevsek, J Redondo,  P Ullio, and A Urbano 
for enlightening conversations, and G Raffelt for 
uncountable useful  discussions and suggestions. 
The project was
performed in the context of the Lyon Institute of Origins, 
grant ANR-10-LABX-66.


\begin{thebibliography}{222222}

\bibitem{rev} for a review, see {\it e.g.}
J.~E.~Kim,
  ``Light Pseudoscalars, Particle Physics and Cosmology,''
  Phys.\ Rept.\  {\bf 150} (1987) 1.

\bibitem{PQ}
  R.~D.~Peccei and H.~R.~Quinn,
  ``CP Conservation in the Presence of Instantons,''
  Phys.\ Rev.\ Lett.\  {\bf 38} (1977) 1440.


  R.~D.~Peccei and H.~R.~Quinn,
  ``Constraints Imposed by CP Conservation in the Presence of Instantons,''
  Phys.\ Rev.\ D {\bf 16} (1977) 1791.





\bibitem{revggr}
 G.~G.~Raffelt,
 ``Astrophysical methods to constrain axions and other novel particle phenomena,''
  Phys.\ Rept.\  {\bf 198} (1990) 1.


 G.~G.~Raffelt,
  ``Stars as laboratories for fundamental physics: The astrophysics of neutrinos, axions, and other weakly interacting particles,''
  Chicago, USA: Univ. Pr. (1996) 664 pp. Or reference \cite{revSred}.

 G.~G.~Raffelt, talk at The Workshop on Baryon and Lepton Number Violation,
Heidelberg, 2013, 
http://www.mpi-hd.mpg.de/BLV2013/

\bibitem{DFS}
  M.~Dine, W.~Fischler and M.~Srednicki,
  ``A Simple Solution to the Strong CP Problem with a Harmless Axion,''
  Phys.\ Lett.\ B {\bf 104} (1981) 199.
 J.~E.~Kim,
  ``Weak Interaction Singlet and Strong CP Invariance,''
  Phys.\ Rev.\ Lett.\  {\bf 43} (1979) 103.



\bibitem{russes}
  M.~A.~Shifman, A.~I.~Vainshtein and V.~I.~Zakharov,
  ``Can Confinement Ensure Natural CP Invariance of Strong Interactions?,''
  Nucl.\ Phys.\ B {\bf 166} (1980) 493.


 A.~R.~Zhitnitsky,
  ``On Possible Suppression of the Axion Hadron Interactions. (In Russian),''
  Sov.\ J.\ Nucl.\ Phys.\  {\bf 31} (1980) 260
   [Yad.\ Fiz.\  {\bf 31} (1980) 497].



\bibitem{DineFischler}
  M.~Dine and W.~Fischler,
  ``The Not So Harmless Axion,''
  Phys.\ Lett.\ B {\bf 120} (1983) 137.



 L.~F.~Abbott and P.~Sikivie,
  ``A Cosmological Bound on the Invisible Axion,''
  Phys.\ Lett.\ B {\bf 120} (1983) 133.

\bibitem{PWW}
 J.~Preskill, M.~B.~Wise and F.~Wilczek,
  ``Cosmology of the Invisible Axion,''
  Phys.\ Lett.\ B {\bf 120} (1983) 127.


\bibitem{autres}
  M.~Kawasaki, K.~Saikawa and T.~Sekiguchi,
  ``Axion dark matter from topological defects,''
  arXiv:1412.0789 [hep-ph].

\bibitem{strings}
  T.~Hiramatsu, M.~Kawasaki, T.~Sekiguchi, M.~Yamaguchi and J.~'i.~Yokoyama,
  ``Improved estimation of radiated axions from cosmological axionic strings,''
  Phys.\ Rev.\ D {\bf 83} (2011) 123531
  [arXiv:1012.5502 [hep-ph]].



  M.~Yamaguchi, M.~Kawasaki and J.~'i.~Yokoyama,
  ``Evolution of axionic strings and spectrum of axions radiated from them,''
  Phys.\ Rev.\ Lett.\  {\bf 82} (1999) 4578
  [hep-ph/9811311].








\bibitem{DavisShellard}
  R.~L.~Davis and E.~P.~S.~Shellard,
  ``Do Axions Need Inflation?,''
  Nucl.\ Phys.\ B {\bf 324} (1989) 167.




\bibitem{DM}
  G.~Bertone, D.~Hooper and J.~Silk,
  ``Particle dark matter: Evidence, candidates and constraints,''
  Phys.\ Rept.\  {\bf 405} (2005) 279
  [hep-ph/0404175].
 {{Jungman}, G. and {Kamionkowski}, M. and {Griest}, K.},
"{Supersymmetric dark matter}",
  Phys.\ Rept.\  {\bf 267} (1996) 195
  [hep-ph/9506380].



\bibitem{ADMX}
 G.~Carosi [ADMX Collaboration],
  ``Searching for old (and new) light bosons with the axion dark matter experiment (ADMX),''
  AIP Conf.\ Proc.\  {\bf 1441} (2012) 494.

\bibitem{EMR}
  D.~Espriu, F.~Mescia and A.~Renau,
  ``Axions and high-energy cosmic rays: Can the relic axion density be measured?,''
  JCAP {\bf 1108} (2011) 002
  [arXiv:1010.2589 [hep-ph]].


\bibitem{axion@WIMP}
 E.~Armengaud, Q.~Arnaud, C.~Augier, A.~Benoit, A.~Benoit, L.~Berg\'e, 
T.~Bergmann and J.~Bl\"umer {\it et al.},
  ``Axion searches with the EDELWEISS-II experiment,''
  JCAP {\bf 1311} (2013) 067
  [arXiv:1307.1488 [astro-ph.CO]].


\bibitem{EdelCDMS}
  Z.~Ahmed {\it et al.}  [CDMS and EDELWEISS Collaborations],
   ``Combined Limits on WIMPs from the CDMS and EDELWEISS Experiments,''
  Phys.\ Rev.\ D {\bf 84} (2011) 011102
  [arXiv:1105.3377 [astro-ph.CO]].

\bibitem{Xenon}
D.~S.~Akerib {\it et al.}  [LUX Collaboration],
  ``First results from the LUX dark matter experiment at the Sanford Underground Research Facility,''
  arXiv:1310.8214 [astro-ph.CO].
 E.~Aprile {\it et al.}  [XENON100 Collaboration],
  ``Dark Matter Results from 225 Live Days of XENON100 Data,''
  Phys.\ Rev.\ Lett.\  {\bf 109} (2012) 181301
  [arXiv:1207.5988 [astro-ph.CO]].
 

\bibitem{SY1}
  P.~Sikivie and Q.~Yang,
  ``Bose-Einstein Condensation of Dark Matter Axions,''
  Phys.\ Rev.\ Lett.\  {\bf 103} (2009) 111301
  [arXiv:0901.1106 [hep-ph]].

\bibitem{ESTY}
  O.~Erken, P.~Sikivie, H.~Tam and Q.~Yang,
  ``Cosmic axion thermalization,''
  Phys.\ Rev.\ D {\bf 85} (2012) 063520
  [arXiv:1111.1157 [astro-ph.CO]].





\bibitem{SY}
  K.~'i.~Saikawa and M.~Yamaguchi,
  ``Evolution and thermalization of dark matter axions in the condensed regime,''
  Phys.\ Rev.\ D {\bf 87} (2013) 085010
  [arXiv:1210.7080 [hep-ph]].


\bibitem{SYetal2}
T.~Noumi, K.~Saikawa, R.~Sato and M.~Yamaguchi,
  ``Effective gravitational interactions of dark matter axions,''
  Phys.\ Rev.\ D {\bf 89} (2014) 065012
  [arXiv:1310.0167 [hep-ph]].



\bibitem{DE}
  S.~Davidson and M.~Elmer,
  ``Bose Einstein condensation of the classical axion field in cosmology?,''
  JCAP {\bf 1312} (2013) 034
  [arXiv:1307.8024].

\bibitem{BJ}
  J.~Berges and J.~Jaeckel,
  ``Far from equilibrium dynamics of Bose-Einstein condensation for Axion Dark Matter,''
  arXiv:1402.4776 [hep-ph].

\bibitem{RDS}
  T.~Rindler-Daller and P.~R.~Shapiro,
  ``Finding new signature effects on galactic dynamics to constrain Bose-Einstein-condensed cold dark matter,''
  arXiv:1209.1835 [astro-ph.CO].
 {\it ibid.}
  ``Angular Momentum and Vortex Formation in Bose-Einstein-Condensed Cold Dark Matter Haloes,''
 Mon.\ Not.\ Roy.\ Astron.\ Soc.\  {\bf 422} (2012) 135
  [arXiv:1106.1256 [astro-ph.CO]].
 {\it ibid.}
  ``Vortices and Angular Momentum in Bose-Einstein-Condensed Cold Dark Matter Halos,'' ASP Conf.\ Ser.\  {\bf 432} (2010) 244
  [arXiv:0912.2897 [astro-ph.CO]].



\bibitem{BS13}
  N.~Banik and P.~Sikivie,
  ``Axions and the Galactic Angular Momentum Distribution,''
 Phys.\ Rev.\ D {\bf 88} (2013) 123517
  [arXiv:1307.3547].



\bibitem{Peebles}
 P.~J.~E.~Peebles,
  ``Fluid dark matter,''
  astro-ph/0002495.
 P.~J.~E.~Peebles,
  ``Dynamics of a dark matter field with a quartic selfinteraction potential,''
  Phys.\ Rev.\ D {\bf 62} (2000) 023502
  [astro-ph/9910350].

\bibitem{WantzShellard}
  O.~Wantz and E.~P.~S.~Shellard,
  ``Axion Cosmology Revisited,''
  Phys.\ Rev.\ D {\bf 82} (2010) 123508
  [arXiv:0910.1066 [astro-ph.CO]].



\bibitem{PDG}
  J.~Beringer {\it et al.}  [Particle Data Group Collaboration],
  ``Review of Particle Physics (RPP),''
  Phys.\ Rev.\ D {\bf 86} (2012) 010001.


\bibitem{WISP}
  J.~Jaeckel and A.~Ringwald,
  ``The Low-Energy Frontier of Particle Physics,''
  Ann.\ Rev.\ Nucl.\ Part.\ Sci.\  {\bf 60} (2010) 405
  [arXiv:1002.0329 [hep-ph]].


\bibitem{BICEP}
 P.~A.~R.~Ade {\it et al.}  [BICEP2 Collaboration],
  ``BICEP2 I: Detection Of B-mode Polarization at Degree Angular Scales,''
  Phys.\ Rev.\ Lett.\  {\bf 112} (2014) 241101
  [arXiv:1403.3985 [astro-ph.CO]].


\bibitem{BGBL}
  M.~Beltran, J.~Garcia-Bellido and J.~Lesgourgues,
  ``Isocurvature bounds on axions revisited,''
  Phys.\ Rev.\ D {\bf 75} (2007) 103507
  [hep-ph/0606107].

 \bibitem{PLANCK}
 P.~A.~R.~Ade {\it et al.}  [Planck Collaboration],
  ``Planck 2013 results. XVI. Cosmological parameters,''
 Astron.\ Astrophys.\  {\bf 571} (2014) A16
  [arXiv:1303.5076 [astro-ph.CO]].



\bibitem{aCDM}
 M.~S.~Turner,
  ``Coherent Scalar Field Oscillations in an Expanding Universe,''
  Phys.\ Rev.\ D {\bf 28} (1983) 1243.

\bibitem{linearaxions} 

 M.~Khlopov, B.~A.~Malomed and I.~.B.~Zeldovich,
  ``Gravitational instability of scalar fields and formation of primordial black holes,''
  Mon.\ Not.\ Roy.\ Astron.\ Soc.\  {\bf 215} (1985) 575.


 J.~-c.~Hwang,
  ``Roles of a coherent scalar field on the evolution of cosmic structures,''
  Phys.\ Lett.\ B {\bf 401} (1997) 241
  [astro-ph/9610042].

\bibitem{HwangNoh}
  J.~-c.~Hwang and H.~Noh,
  ``Axion as a Cold Dark Matter candidate,''
  Phys.\ Lett.\ B {\bf 680} (2009) 1
  [arXiv:0902.4738 [astro-ph.CO]].

\bibitem{ratra}
  B.~Ratra,
  ``Expressions for linearized perturbations in a massive scalar field dominated cosmological model,''
  Phys.\ Rev.\ D {\bf 44} (1991) 352.

\bibitem{NambuSasaki}
  Y.~Nambu and M.~Sasaki,
  ``Quantum Treatment Of Cosmological Axion Perturbations,''
  Phys.\ Rev.\ D {\bf 42} (1990) 3918.



\bibitem{mini}
  C.~J.~Hogan and M.~J.~Rees,
  ``Axion Miniclusters,''
  Phys.\ Lett.\ B {\bf 205} (1988) 228.

\bibitem{hbar}
  S.~J.~Brodsky and P.~Hoyer,
  ``The $\hbar$ Expansion in Quantum Field Theory,''
  Phys.\ Rev.\ D {\bf 83} (2011) 045026
  [arXiv:1009.2313 [hep-ph]].



  B.~R.~Holstein and J.~F.~Donoghue,
  ``Classical physics and quantum loops,''
  Phys.\ Rev.\ Lett.\  {\bf 93} (2004) 201602
  [hep-th/0405239].

  N.~E.~ Bjerrum-Bohr, J.~F.~Donoghue and B.~R.~Holstein,
  ``Quantum gravitational corrections to the nonrelativistic scattering potential of two masses,''
  Phys.\ Rev.\ D {\bf 67} (2003) 084033
   [Erratum-ibid.\ D {\bf 71} (2005) 069903]
  [hep-th/0211072].
 

  C.~Montonen and D.~I.~Olive,
  ``Magnetic Monopoles as Gauge Particles?,''
  Phys.\ Lett.\ B {\bf 72} (1977) 117.





\bibitem{I+Z}
C Cohen-Tannoudji, B Diu, F Lalo\"e,
M\'echanique Quantique, vol 1, chapter 5.G, Hermann, 1977.\\
C Itzykson, J.B. Zuber, ``Quantum Field Theory'', McGraw-Hill, New York, USA.

\bibitem{AB}
  G.~Aarts and J.~Berges,
  ``Classical aspects of quantum fields far from equilibrium,''
  Phys.\ Rev.\ Lett.\  {\bf 88} (2002) 041603
  [hep-ph/0107129].


\bibitem{Mot}
  F.~Cooper and E.~Mottola,
  ``Initial Value Problems in Quantum Field Theory in the 
Large N Approximation,''
  Phys.\ Rev.\ D {\bf 36} (1987) 3114.



\bibitem{Schwinger}
  J.~S.~Schwinger,
  ``Brownian motion of a quantum oscillator,''
  J.\ Math.\ Phys.\  {\bf 2} (1961) 407.
 P.~M.~Bakshi and K.~T.~Mahanthappa,
  ``Expectation value formalism in quantum field theory ''
  J.\ Math.\ Phys.\  {\bf 4} (1963) 1.
 J.\ Math.\ Phys.\  {\bf 4} (1963) 12.
L. V. Keldysh,  Sov.\ Phys.\ JETP  20, (1965)  1018.
 K.~-c.~Chou, Z.~-b.~Su, B.~-l.~Hao and L.~Yu,
  ``Equilibrium and Nonequilibrium Formalisms Made Unified,''
  Phys.\ Rept.\  {\bf 118} (1985) 1.



\bibitem{CJT}
  J.~M.~Cornwall, R.~Jackiw and E.~Tomboulis,
  ``Effective Action for Composite Operators,''
  Phys.\ Rev.\ D {\bf 10} (1974) 2428.
 
\bibitem{RamseyHu}
  S.~A.~Ramsey and B.~L.~Hu,
  ``O(N) quantum fields in curved space-time,''
  Phys.\ Rev.\ D {\bf 56} (1997) 661
  [gr-qc/9706001].


\bibitem{Bog}
N.~N.~Bogoliubov,
J. Phys. \ (Moscow)  {\bf 11} (1947) 23.

\bibitem{KapustaBEC}
  J.~I.~Kapusta,
  ``Bose-Einstein Condensation, Spontaneous Symmetry Breaking, and Gauge Theories,''
  Phys.\ Rev.\ D {\bf 24} (1981) 426.





\bibitem{HW}
  H.~E.~Haber and H.~A.~Weldon,
  ``Finite Temperature Symmetry Breaking as Bose-Einstein Condensation,''
  Phys.\ Rev.\ D {\bf 25} (1982) 502.




\bibitem{Leggett}
  A.~J.~Leggett,
  ``Bose-Einstein condensation in the alkali gases: Some fundamental concepts,''
  Rev.\ Mod.\ Phys.\  {\bf 73} (2001) 307
   [Erratum-ibid.\  {\bf 75} (2003) 1083].


\bibitem{PitRev}
F Dalfovo, S Giorgini, L.P. Pitaevskii, S. Stingari,
``Theory of Bose-Einstein condensation in trapped gases'',
Rev. Mod Phys. {\bf 71} (1999), 463.


\bibitem{efstathiou}
  G.~Efstathiou,
  ``Cosmological perturbations,''
  In *Edinburgh 1989, Proceedings, Physics of the early universe* 361-463.


\bibitem{Wigner}
 E.~Calzetta and B.~L.~Hu,
  ``Nonequilibrium Quantum Fields: Closed Time Path Effective Action, Wigner Function and Boltzmann Equation,''
  Phys.\ Rev.\ D {\bf 37} (1988) 2878.
C. Cardall,
  Phys.\ Rev.\ D {\bf 78} (2008)  085017.




\bibitem{MB}
  C.~-P.~Ma and E.~Bertschinger,
  ``Cosmological perturbation theory in the synchronous and conformal Newtonian gauges,''
  Astrophys.\ J.\  {\bf 455} (1995) 7
  [astro-ph/9506072].



\bibitem{planck}
  S.~M.~Barr and D.~Seckel,
  ``Planck scale corrections to axion models,''
  Phys.\ Rev.\ D {\bf 46} (1992) 539.
 R.~Kallosh, A.~D.~Linde, D.~A.~Linde and L.~Susskind,
  ``Gravity and global symmetries,''
  Phys.\ Rev.\ D {\bf 52} (1995) 912
  [hep-th/9502069].
 G.~Dvali, S.~Folkerts and A.~Franca,
  ``On How Neutrino Protects the Axion,''
  arXiv:1312.7273 [hep-th].
 


\bibitem{DeWitt}
  B.~S.~DeWitt,
  ``Quantum Theory of Gravity. 3. Applications of the Covariant Theory,''
  Phys.\ Rev.\  {\bf 162} (1967) 1239.
 


\bibitem{NZG} T. Nikuni, E. Zaremba, and A. Griffin,
``Two-Fluid Dynamics for a Bose-Einstein Condensate 
out of Local Equilibrium with the Noncondensate ``,
Phys. \ Rev.\  Lett. 83 (1999), 10. 
[cond-mat/9812320]

\bibitem{revSred}
  M.~Srednicki,
  ``Axion Couplings to Matter. 1. CP Conserving Parts,''
  Nucl.\ Phys.\ B {\bf 260} (1985) 689.


\end{thebibliography}
\end{document}